\documentclass{article}

\usepackage{graphicx}
\usepackage{amsmath}
\usepackage{amssymb}
\usepackage{epstopdf}
\usepackage{mathrsfs}
\usepackage{mathtools}

\title{Bell's Conspiracy, Schr\"{o}dinger's Black Cat and Global Invariant Sets}

\author{T.N.Palmer \\ Department of Physics, University of Oxford}

\begin{document}

\bibliographystyle{plain}

\maketitle

\makeatletter
\newcommand\be{\@ifstar{\[}{\begin{equation}}}
\newcommand\ee{\@ifstar{\]}{\end{equation}}}
\newcommand\bp{\begin{pmatrix}}
\newcommand\ep{\end{pmatrix}}
\newcommand\ua{\uparrow}
\newcommand\da{\downarrow}
\makeatother 

\begin{abstract}
A locally causal hidden-variable theory of quantum physics need not be constrained by the Bell inequalities if this theory also partially violates the measurement independence condition. However, such violation can appear unphysical, implying implausible conspiratorial correlations between the hidden-variables of particles being measured and earlier determinants of instrumental settings. A novel physically plausible explanation for such correlations is proposed, based on the hypothesis that states of physical reality lie precisely on a non-computational measure-zero dynamically invariant set in the state space of the universe: the Cosmological Invariant Set Postulate. To illustrate the relevance of the concept of a global invariant set, a simple analogy is considered where a massive object is propelled into a black hole depending on the decay of a radioactive atom. It is claimed that a locally causal hidden-variable theory constrained by the Cosmological Invariant Set Postulate can violate the CHSH inequality without being conspiratorial, superdeterministic, fine-tuned or retrocausal, and the theory readily accommodates the classical compatibilist notion of (experimenter) free will.
 \end{abstract}

\newpage

\section{Introduction}
\label{introduction}

It is well known that a locally causal hidden-variable theory of quantum physics need not be constrained by the Bell inequalities if this theory also violates the measurement independence condition
\be
\label{mip}
\rho(\lambda| a, b)= \rho (\lambda| a', b')
\ee
for all $a, a', b, b'$. Here $\rho(\lambda| a, b)$ denotes a probability density function over some hidden-variable $\lambda$, conditional on the orientations of Alice and Bob's measuring apparatuses, represented by the points $a$ and $a'$ (for Alice), and $b$ and $b'$ (for Bob), on the unit sphere. In fact, it may only be necessary to violate (\ref{mip}) partially \cite{Hall:2010} \cite{Hall:2011}, suggesting that this might be a fruitful area to explore in order to create viable realistic causal theories of fundamental physics. However, an important objection to any proposed violation of measurement independence is that it can give rise to correlations that seemingly have no basis in physical theory, suggesting some implausible conspiracy operating in the universe at large. 

Consider the example proposed by Bell himself \cite{Bell} where a measurement orientation is set by some pseudo-random number generator (PRNG). The output of the PRNG is determined by, and depends sensitively on, some input variable. For example, the value of the most significant digit of the output variable may depend on the value of the millionth digit of the input variable. In Bell's own words:
\begin{quote}
\ldots this particular piece of information is unlikely to be the vital piece for any distinctively different purpose, i.e. is otherwise rather useless.
\end{quote}
Hence, if we seek to violate measurement independence, even partially, then we must explain how the value of this millionth digit can somehow correlated with the value of the hidden variable of the particles being measured. This becomes especially problematic when we realise that the PRNG may be run months (or indeed billions of years, see \cite{Gallicchio:2014}) before the particles being measured are actually produced as part of the experimental procedure. One proposal to explain such correlations is retrocausality \cite{Price:1997}.

Here, an alternative and novel way to understand these correlations is proposed, based on the concept of global invariant sets in state space, and fractal invariant sets in particular. Importantly, local properties of such sets cannot be determined by finite algorithm - they are formally non-computational. Whilst generic to a broad class of nonlinear dynamical system, fractal invariant sets are not especially familiar concepts to many quantum physicists and philosophers. Hence, in Section \ref{cat}, to illustrate the power of the global invariant set in providing a new perspective on this long-studied problem, we consider a simple analogy: a gedanken experiment in which, depending on whether a radioactive atom decays, a massive object is fired into a black hole. We consider the temporal correlation between the size of the black hole at $t_0$, as determined by the radius of its event horizon $\mathscr H ^+ (t_0)$, and the (supposed) hidden variable of the radioactive atom at $t_1$. The existence of such a correlation, even when $t_0 \ll t_1$, is easy to understand (without conspiracy or retrocausality) if we take into account the fact that $\mathscr H ^+$ is a globally defined invariant subset of space time. In particular, like the fractal invariant set, $\mathscr H^+(t_0)$'s local properties cannot be defined from the neighbouring space-time geometry near $t_0$. 

Two properties of fractal attracting invariant sets in state space are discussed in Section \ref{cantor}: their non-computability, and the notion that such sets are, in a well defined sense, `large' when looked at from the inside, yet `small' when looked at from the outside. The key thesis which underpins the subsequent analysis in this paper is then proposed: that states of physical reality lie precisely on a (fractal) invariant set in the state space of the universe - the Cosmological Invariant Set Postulate. In Section \ref{CHSH} it is shown how a development of these ideas, referred to as Invariant Set Theory, can nullify the CHSH version of the Bell Theorem, determinism and causality notwithstanding. In Section \ref{conspiracy} it is shown that the proposed nullification suffers from none of the familiar objections: it is not conspiratorial, it does not need retrocausality, the theory is not superdeterministic or fine tuned, it is robust under small perturbations, and experimenters are not prevented from measuring particle spins relative to any direction they may wish to measure. Despite this, the Cosmological Invariant Set Postulate is manifestly not classical. Based on the discussion in this paper, in Section \ref{conclusions}, some remarks are made about the relevance of the complex Hilbert Space as a description of physical reality. 

\section {Schr\"{o}dinger's Black Cat}
\label{cat}

Consider the following simple gedanken experiment. If a given radioactive atom decays (within a certain period of time) then a massive object - which could in principle be a cat in a box - is propelled into a black hole. If the radioactive atom does not decay, the massive object remains some distance from the black hole in question. 

\begin{figure}
\centering
\includegraphics[scale=0.3]{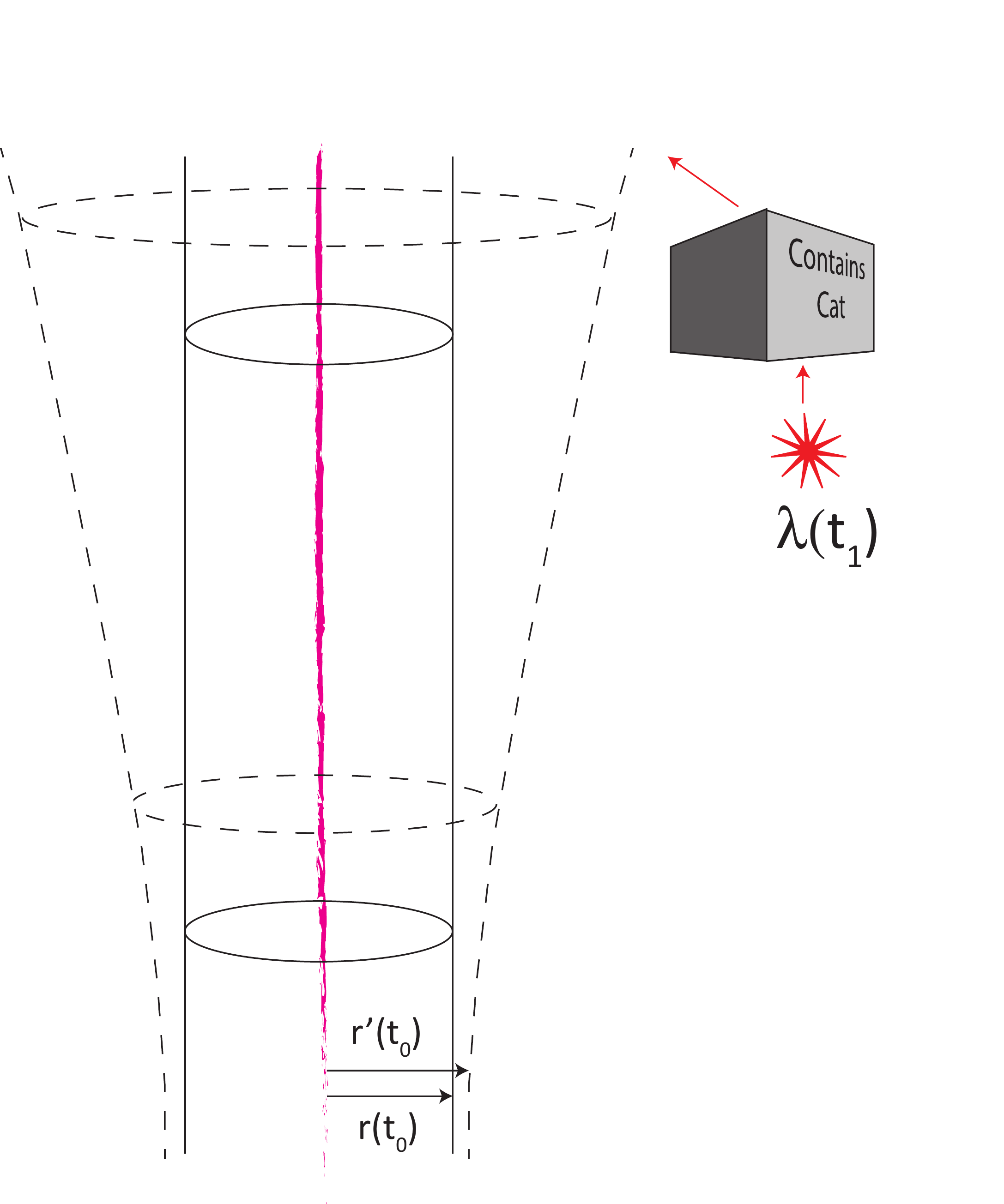}
\caption{A massive object is propelled into a black hole if a radioactive atom decays within a certain period of time around $t_1$. Assuming some underpinning hidden variable theory, the size of the black hole at $t_0< t_1$ depends on whether the radioactive atom decays and hence on the value of the atom's hidden variables.  If the size of the black hole could be estimated from the space-time geometry in the neighbourhood of the black hole at $t_0$, then this dependency would appear to imply either some implausible conspiracy or some form of retrocausality. However, since the size of the black hole is determined by the distance of the event horizon $\mathscr H ^+$ from the centre of the black hole, and since $\mathscr H^+$ is itself defined by a global space-time condition (the boundary of null surfaces that extend to $\mathscr I ^+$), neither conspiracy or retrocausality is needed to explain this dependency.}
\label{fig:blackhole}
\end{figure}

The mass and hence size of the black hole are therefore dependent on the decay of the radioactive atom. As shown in Fig \ref{fig:blackhole}, let the size of the black hole in the case where the radioactive atom does not decay be equal to the radius $r(t)$ of the event horizon $\mathscr H ^+$, where $t$ labels a family of spacelike hypersurfaces which intersect $\mathscr H ^+$. Fig \ref{fig:blackhole} shows a second null surface lying at a radius $r'(t)> r(t)$. If the radioactive atom does decay at time $t_1$ then whilst for $t<t_1$ it appears that this null surface will escape to future null infinity, $\mathscr I ^+$, in fact after the object has fallen into the black hole, increasing its mass, this null surface becomes trapped. The second null surface can be assumed to correspond to the event horizon of the black hole in the case where the atom decays. Hence at $t_0<t_1$ the size of the black hole is bigger if the atom decays at $t_1$ than if it doesn't. If the mass of the `cat+box' is large enough, then even at times $t_0 \ll t_1$, it is possible that $r'(t_0) \gg r(t_0)$. 
In the conventional language of quantum theory (assuming the whole system is unobserved) the black hole at the earlier time $t_0$ must be considered a linear superposition of black holes of size $r$ and $r'$. Here we pursue the possibility that quantum physics is indeed underpinned by some causal hidden-variable theory. Hence, let the decay of the radioactive atom at $t_1$ be determined by the value of some hidden variable $\lambda$. Then, according to the discussion above, the size of the black hole at $t_0 < t_1$ is both well-defined (i.e. not a superposition of sizes) and correlated with $\lambda$ at $t_1$. Instead of a radioactive atom decay, the release of the box could be determined by an experimenter. In this case, the size of the black hole at $t_0$ will be the result of experimenter decision at $t_1$. 

A physicist who believed that the size of the black hole at $t_0$ can be determined from the space-time geometry in some neighbourhood of $\mathscr H^+$ at $t_0$ would be faced with a dilemma. Somehow, the black hole would have to `know' to be the right size at $t_0$ in order to anticipate the decay or non-decay of the atom, or alternatively the experimenter decision, at $t_1$. The apparent dilemma is compounded by the fact that the photons which propagate on $\mathscr H^+$ never interact with the radioactive atom (or experimenter) at $t_1$. To avoid requiring some implausible conspiracy, the physicist may conclude that there may be a retrocausal effect by which information, associated with the decay of the radioactive atom, propagates back in time causing the black hole event horizon to expand at earlier times.  

However, in this case there is a simple explanation for the correlation which requires neither conspiracy nor retrocausality. The explanation lies in the fact that $\mathscr H^+$ is defined by a global property of space-time, it being the boundary of null surfaces that escape to $\mathscr I^+$. Consistent with this, the location $r(t_0)$ of $\mathscr H^+$ cannot be determined by the geometry of space time in the neighbourhood of $\mathscr H^+$ at $t_0$. Hence, the correlation is explainable without conspiracy or retrocausality. 

\section{Global Invariant Sets in State Space}
\label{cantor}

Instead of the global geometric constructions in space-time, we now consider global geometric constructions in state space. 

Consider a dynamical system $\mathcal D: X_{t_0} \rightarrow X_{t_0+t}$, defined, say, by a set of differential equations $\dot X=F[X]$. An attracting invariant set $I_{\mathcal D}$ is a closed subset of the state space of $\mathcal D$ with the properties (see e.g. \cite{Strogatz}):

\begin{itemize}
\item If $X_{t_0} \in {I}_{\mathcal D}$ then $X_{t_0+t} \in {I}_{\mathcal D}$ for all $t$.
\item $I_{\mathcal D}$ attracts an open set of initial conditions. That is, there is an open neighbourhood $\mathcal U$ containing ${I}_{\mathcal D}$ such that if $X(0) \in \mathcal U$, then the distance from $X(t)$ to $I _ {\mathcal D}$ tends to zero as $t \rightarrow \infty$. The largest such $\mathcal U$ is called the basin of attraction of $\mathcal {I}_{\mathcal D}$. 
\item There is no proper subset of ${I}_{\mathcal D}$ having the first two properties. 

\end{itemize}
Just as $\mathscr H^+$ is a subset of space-time defined globally, so $I_{\mathcal D}$ is a subset of state space defined globally. One can readily give simple examples of dynamical systems with attracting invariant sets. Consider for example the dynamical system whose evolution equations are 
\begin{equation}
\dot r=r(1-r^2) \;\;\;\; \dot \theta=1
\end{equation}
in polar coordinates $(r,\theta)$. It is easily shown that all trajectories spiral asymptotically towards a limit cycle at $r=1$. The attracting invariant set is the subset of points where $r=1$. More generally, there is a generic  class of nonlinear dynamical systems where the attracting invariant sets are not such simple subsets, but are fractal eg the Lorenz \cite{Lorenz:1963} attractor $I_L$, derived from the equations
\begin{eqnarray}
\label{lorenz}
\dot X &=& -\sigma X + \sigma Y \nonumber \\
\dot Y &=& -XZ+rX-Y \nonumber \\
\dot Z &=&  \;\;\;XY -bZ 
\end{eqnarray}
where $\sigma, r, b >0$. For such dynamical systems there is no formula defining the invariant set. Consider then the question of determining whether a given point in state space lies in $I_L$. One may start with a state known to be in the basin of attraction of $I_L$. However, one would have to integrate the evolution equations for an infinite time before one could be sure that the state had evolved onto the invariant set. One would then integrate for a further time, comparing the evolved state with the chosen point in state space, and stopping if the two were identical. This cannot be achieved by a finite algorithm and one must therefore conclude that the fractal invariant set is not computational. More formally, Blum et al \cite{Blum} have shown that for decision processes such as this to be solved in finite time, the invariant set must have integral Hausdorff dimension. By definition, fractal attracting invariant sets do not have integral Hausdorff dimension. Dube \cite{Dube:1993} has shown that some of the classic problems in computation theory that cannot be solved by algorithm, can be recast in terms of fractal geometry. For example, the Post Correspondence Problem is equivalent to that of determining whether a given line intersects the attracting invariant set of an Iterative Function System. 

A fractal invariant set can locally be considered the Cartesian product $\mathbb R \times \mathcal C$ where $\mathcal C$ denotes some (multidimensional) Cantor set. The simplest example of a Cantor Set is the ternary set $\mathcal T$, based on the intersection of all iterates $\mathcal T_k$, where $\mathcal T_{k}$ comprises two copies of $\mathcal T_{k-1}$, each copy reduced by a factor of $1/3$. Generalisations of $\mathcal T$ are needed to define invariant set geometries which exhibit quantum probability structure \cite{Palmer:2014} \cite{Palmer:2015b}. Key properties shared by all Cantor sets are the following:
\begin{itemize}
\item Relative to the full measure of the Euclidean space in which they are embedded, Cantor Sets have measure zero (i.e. are `small' when looked at from the outside).
\item Despite this Cantor sets have the cardinality of the continuum (i.e. are `large' when looked at from the inside.
\end{itemize}
For example, elements of $\mathcal T$ can be represented by the numbers $0 \le r \le 1$ whose base-3 expansion contain no digit `1'. The probability that a randomly chosen number on the interval $[0,1]$ has such a base-3 expansion is equal to zero. However, if we take the base-3 representation of a point in $\mathcal T$ and replace all occcurrences of `2' with the digit `1', and then interpret that number as a real number in binary form, it can be seen that there are as many points in $\mathcal T$ are there are points in $[0,1]$. The ternary set $\mathcal T$ can be readily generalised, firstly be extending into multiple state-space dimensions (e.g. the Menger Sponge) and secondly by considering restrictions on base-$N$ rather than base-$3$ numbers, where $N>3$. In this way, for large enough $N$ it is possible to construct multidimensional fractals which, whilst still measure zero in their embedding space, do not 'appear' at all gappy (or lacunar \cite{Mandelbrot}). These are relevant in the discussion below. It should also be noted that a Cantor Set is an example of what, mathematically, is referred to as a 'perfect set". That is to say, given any point of the Cantor Set and any neighbourhood of this point in the embedding space contains another point of the Cantor Set - i.e. the Cantor Set has no isolated points. This is relevant in the discussion about `superdeterminism' below. 

There is an important technique for reconstructing an attracting invariant set using data from a (low-dimensional) dynamical system. It is based on the notion that the dynamics in the full state space can be reconstructed from measurements of just a single component of the state space \cite{Takens:1981}: The Takens Embedding Theorem. The method is based on time delays. This fact has important conceptual implications discussed later. 

\begin{figure}
\centering
\includegraphics[scale=0.4] {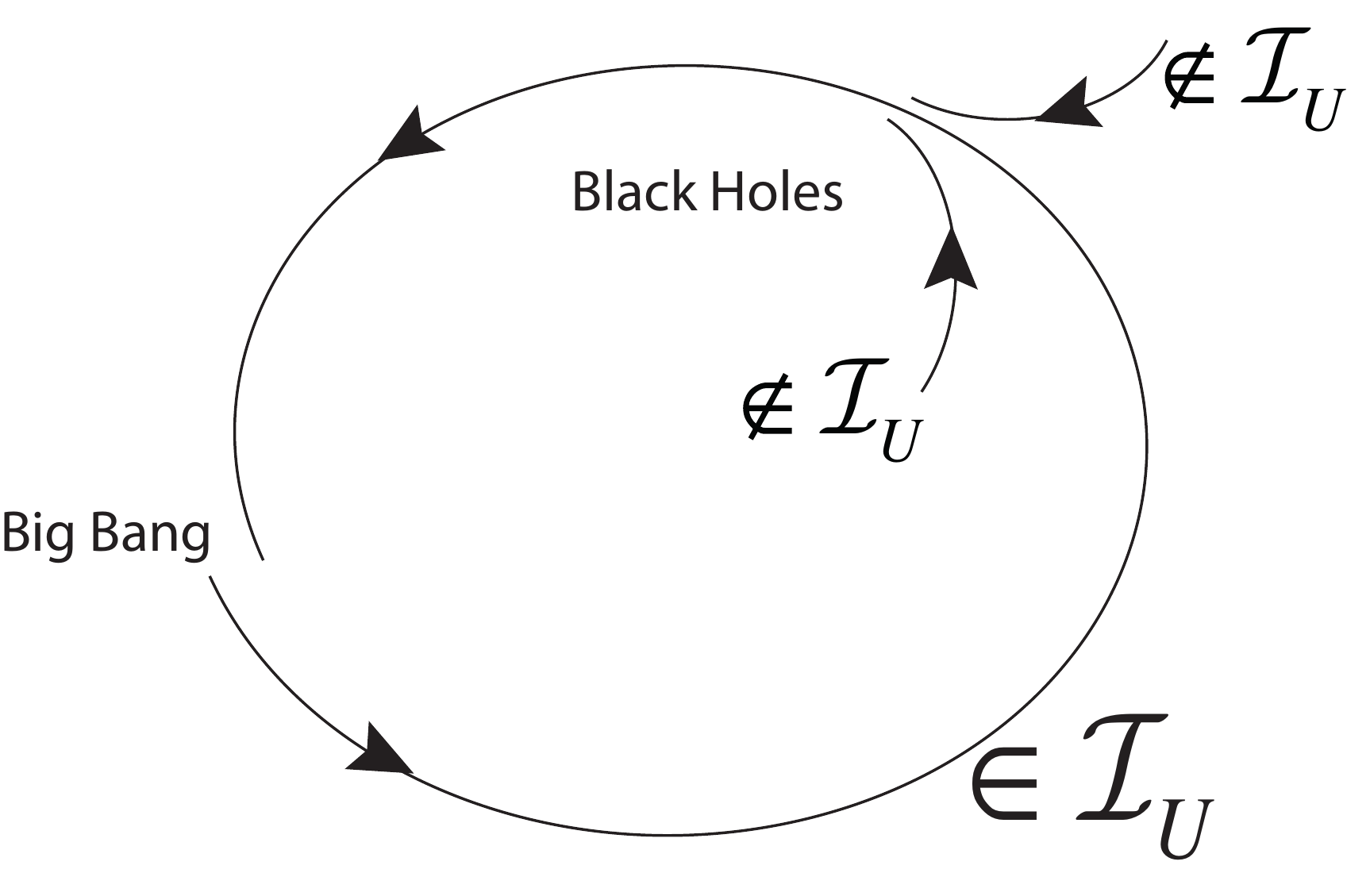}
\caption{A schematic illustration of the evolving trajectory of the universe $U$ in its state space, within a quasi-cyclic cosmology. The Cosmological Invariant Set Postulate states that these trajectories are evolving on a measure-zero fractal attracting invariant subset $I_U$ which defines $U$'s dynamical evolution. In regions of state space containing black holes, the black hole no-hair theorem implies that state space trajectories are convergent (consistent with discussions by Penrose \cite{Penrose:2010}).}
\label{fig:bigbang}
\end{figure}

We apply these concepts to fundamental physics through the Cosmological Invariant Set Postulate \cite{Palmer:2014}: states of physical reality lie precisely on a fractal attracting invariant set $I_U$ of measure zero in the state space of the universe $U$, considered as a dynamical system. A crucial question to ask is: What fundamental physical process could account for the attraction of neighbouring state-space trajectories onto $I_U$? Here we appeal to ideas put forward by Roger Penrose (e.g. \cite{Penrose:2010}), that the black hole no-hair theorem (and consequent quantum black-hole information loss) should imply the existence of regions of state-space trajectory convergence. The idea is illustrated schematically in Fig \ref{fig:bigbang}, which shows the trajectory of the universe in its state space over a cycle of an assumed quasi-cyclic cosmology. By the Cosmological Invariant Set Postulate, this trajectory lies precisely on $I_U$. Shown in Fig \ref{fig:bigbang} are regions of state space (containing black holes) where neighbouring trajectories are converging onto $I_U$. It should be noted that these convergent trajectories do not lie on $I_U$, leading to some important new perspectives on the arrow of time problem that will be addressed elsewhere. 

\section{The Bell Theorem}
\label{CHSH} 

The relevance of the Cosmological Invariant Set Postulate to the Bell Theorem can now be discussed. The key conceptual issues are related to the properties in the two bullets above:
\begin{itemize}
\item $I_U$ is sufficiently small from the perspective of the embedding space that the counterfactual experiments needed to establish the Bell inequality do not lie on $I_U$ and therefore, by postulate, do not correspond to states of physical reality.
\item $I_U$ is sufficiently large that these restrictions do not pose any practical constraints on which experiments experimenters might want to perform. Related to this, the theory which arises out of the Cosmological Invariant Set Postulate cannot be said to be `superdeterministic' or fine tuned in any meaningful sense. 
\end{itemize}
Before discussing these issues in detail, a brief summary of some ongoing technical developments to describe the geometry of $I_U$ using algebraic techniques - something referred to below as `invariant set theory' (see \cite {Palmer:2014} \cite{Palmer:2015b}). 

\subsection{Some Technical Preliminaries} 

As discussed above, locally, $I_U$ can be written as $\mathbb{R} \times \mathcal C$ where $\mathcal C$ is a Cantor Set. That is to say, locally $I_U$ is a Cantor Set of trajectories, each trajectory representing a cosmological space-time. Using the language of symbolic dynamics \cite{Williams} and recognising Schwinger's \cite{Schwinger} symbolic algebraic approach to quantum mechanics, a theoretical framework is developed in \cite{Palmer:2015b} which defines a one-to-one mapping (which, importantly, is not a surjection) between a symbolic representation of fractal trajectory bundles on $I_U$ and the (multi-qubit) complex Hilbert Space of quantum theory. 
 
For one qubit, this correspondence can be written
 \begin{equation}
\label{correspondence}
\mathbf{E}^\alpha_\beta(000\ldots 0)^T \sim
 \cos\frac{\theta}{2}\;|0\rangle + \sin\frac{\theta}{2} e^{i \phi}\;|1\rangle
 \end{equation}
where $\mathbf{E}^\alpha_\beta(000\ldots 0)^T$ is a bit string of length $2^N$ comprising the symbols '0' and '1', $N$ is a large but finite integer from which the fractal dimension of $I_U$ is determined, and the superscript `T' denotes matrix transpose. From the perspective of Invariant Set Theory, the symbols `0' and `1' describe discrete attracting subsets of $I_U$; from a quantum theoretic perspective these symbols describe measurement outcomes\footnote{In \cite{Palmer:2015b} these attracting subsets are considered state-space manifestations of gravitational `clumpiness' - thus linking Invariant Set Theory to descriptions of quantum measurement as a fundamentally gravitational process}. From the perspective of Invariant Set Theory, the bit strings $\mathbf{E}^\alpha_\beta(000\ldots 0)$ describe a bundle of trajectories belonging to a particular fractal iterate, $\mathcal C_k$ of $\mathcal C$; from a quantum theoretic perspective, these bit strings define the sample spaces from which measurement probabilities are defined. In Invariant Set Theory, $\mathbf{E}^\alpha_\beta$ denotes a two parameter family of negation/permutation operators acting on bit strings, with the multiplicative property of quaternions (and hence Pauli Spin matrices). In particular, the frequency of occurrence of the symbol `0' in  $\mathbf{E}^\alpha_\beta(000 \ldots 0)^T$ is equal to $|1-\alpha/2|$. Importantly, $\alpha$ and $\beta$ are only defined if their binary representations can be expressed with less than $N$ bits. In particular, $\mathbf{E}^\alpha_\beta$ is undefined if either $\alpha$ or $\beta$ is irrational. 

These numbers relate to complex Hilbert Space parameters through the relationships
\begin{align}
\label{incom}
\cos^2 \theta/2&= |1-\alpha/2| \nonumber \\
\phi&=\pi\beta/2 
\end{align}
Hence, the frequency of occurrence of the symbol `0' in  $\mathbf{E}^\alpha_\beta(000 \ldots 0)^T$ is equal to $\cos^2\theta/2$, and, for example, the eigenstates $|0\rangle$ and $|1 \rangle$ in quantum theory correspond to the bit strings $\mathbf{E}^0_\beta (000 \ldots 0)=(000 \ldots 0)$ and $\mathbf{E}^2_\beta (000 \ldots 0)=(111 \ldots 1)$ respectively.  The fractal structure of $I_U$ becomes relevant when considering multiple sequential spin measurements. Since $\alpha$ and $\beta$ must be expressible with $N$ bits, so too must $\theta$ and $\phi$. This means that according to Invariant Set Theory, complex Hilbert Space vectors where either $\theta$ or $\phi$ is not describable by $N$ bits (e.g. are irrational) do not correspond to a sample space of trajectories on the invariant set $I_U$, and therefore, by the Cosmological Invariant Set Postulate, do not corresponds to a sample space comprising elements of physical reality. Put another way, such Hilbert Space vectors describe physically unreal counterfactual experiments. 

The relationship (\ref{correspondence}) can be readily extended to $M \ge 1$ qubits, corresponding to $M$ bit strings based on multiple copies of the quaternion operators $\mathbf{E}^\alpha_\beta$. We refer the reader to reference \cite{Palmer:2015b} for details. 

We can now discuss how precisely invariant set theory fails to be constrained by the Bell inequality, and the CHSH \cite{CHSH} version of it in particular (similar arguments apply to an analysis of the original Bell inequalities \cite{Palmer:2014}). 

\subsection{CHSH}

\begin{figure}
\centering
\includegraphics[scale=0.7]{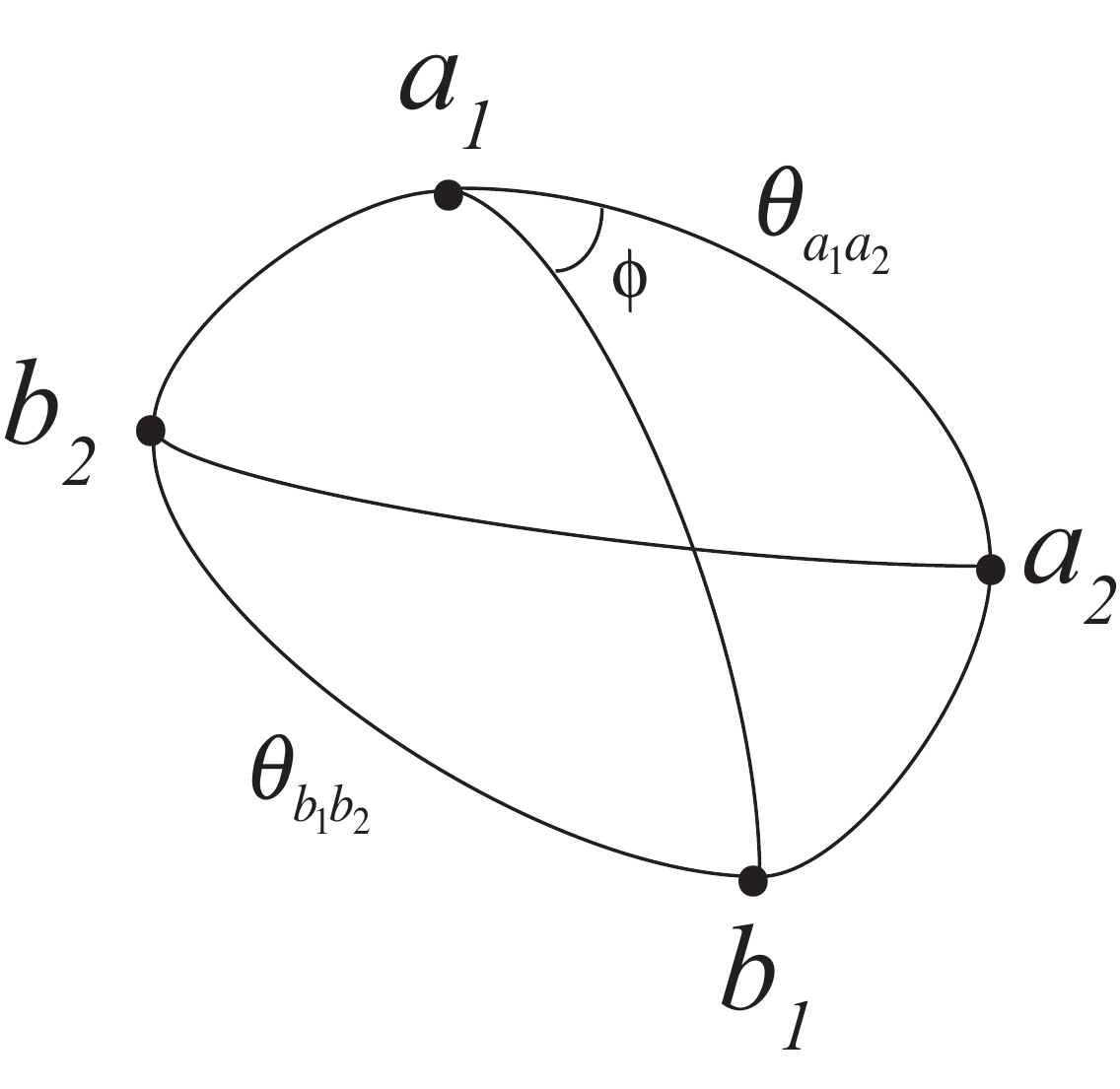}
\caption{$a_1, a_2, b_1, b_2$ are four points on the sphere, representing directions associated with measurement orientations from which the left hand side of the CHSH inequality (\ref{chsh}) can be estimated. We focus on a particular $\lambda \in \Lambda$. The angular length between $a_i$ and $b_j$ be $\theta_{a_i,b_j}$. According to Invariant Set Theory, in order that $\mathrm{Corr}_{\Lambda}(a_i b_j)$ in (\ref{chsh}) be definable, then $\cos \theta_{a_ib_j}$ must be base-2 rational. As discussed in the text $\cos \theta_{a_1a_2}$ and $\cos \theta_{b_1b_2}$ are necessarily base-2 rational. Then, using the cosine rule for spherical triangles, it is impossible for the sides of any of the spherical triangles whose apexes are drawn from $\{a_1, a_2, b_1, b_2\}$ to all have angular lengths whose cosines are base-2 rational. Based on this, the left hand side of the CHSH inequality is not definable from a common sample space of hidden variables, and Invariant Set Theory is therefore not constrained by the CHSH inequality.}
\label{fig:chsh}
\end{figure}

Alice and Bob each have a measuring device which has two orientation settings ($a_1$ and $a_2$ for Alice, and $b_1$ and $b_2$ for Bob) and two 'outputs' say $\pm1$.  Let the orientations $a_1$, $a_2$, $b_1$ and $b_2$ be represented by four points on the unit sphere (see Fig \ref{fig:chsh}). A conventional causal hidden-variable theory is constrained by the CHSH inequality
\begin{equation}
\label{chsh}
|\mathrm{Corr}_{\Lambda}(a_1, b_1) - \mathrm{Corr}_{\Lambda}(a_1, b_2)|+|\mathrm{Corr}_{\Lambda}(a_2, b_1)+\mathrm{Corr}_{\Lambda}(a_2, b_2)| \le 2
\end{equation}
where each correlation is defined by
\be
\label{correlation}
\mathrm{Corr}_{\Lambda}(a_i, b_j)=\sum_{k} A(a_i, \lambda_k) B(b_j, \lambda_k)
\ee
where $A=\pm1$ and $B=\pm1$ denote deterministic (hidden-variable) functions and $\Lambda=\{\lambda_k\}$. As is well known, according to experiment and consistent with quantum theory, with $a_1 \approx 0^\circ$, $a_2 \approx 90^\circ$, $b_1 \approx 45^\circ$, $b_2 \approx 135^\circ$, then the left hand side of (\ref{chsh}) sums to about 2.8, in clear violation of (\ref{chsh}). Does a causal hidden-variable theory constrained by the Cosmological Invariant Set Postulate necessarily obey the CHSH inequality? According to the discussion above, for Invariant Set Theory to necessarily obey the CHSH inequality, then the cosine of the angular distance between any pair of the four points $a_1, a_2, b_1, b_2$ must be describable by $O(N)$ bits. If this is possible, then the sample spaces which generate the correlations $\mathrm{Corr}_{\Lambda}(a_1, b_1), \mathrm{Corr}_{\Lambda}(a_1, b_2), \mathrm{Corr}_{\Lambda}(a_2, b_1)$ and $\mathrm{Corr}_{\Lambda}(a_2, b_2)$ can be associated with bit strings of the type $\{a_1, a_2, \ldots a_{2^N}\}$, where $a_i \in \{0,1\}$, and the symbol $0$ denotes a situation where $A \times B=1$ (Alice and Bob agree) and the symbol $1$ denotes a situation where $A \times B = -1$ (Alice and Bob disagree). 

However, it is not the case that the cosine of the angular distance between all pairs of points $a_1, a_2, b_1, b_2$ are base-2 rational (and hence cannot be describable by $N$ bits). To see this, first of all note that $\cos \theta_{a_1a_2}$ must be describable by $N$ bits (where $\theta_{a_1a_2}$ denotes the relative orientation between $a_1$ and $a_2$). The reason for this is that it is possible for Alice to measure the spin of a particle with her apparatus oriented with respect to $a_1$, and, with the spin now prepared in the $a_1$ direction, to measure the same particle again with the apparatus oriented in the $a_2$ direction. For an experiment where a particle is prepared in the state associated with $a_1$ and measured in the state associated with $a_2$, then according to Invariant Set Theory, $\cos\theta_{a_1a_2}$ must be describable by $N$ bits. Similarly, $\cos \theta_{b_1b_2}$ must be describable by $N$ bits. 

As discussed below, Alice and Bob can be considered, for all practical purposes, free agents. Therefore they can choose, without constraint, any of the following possibilities $\{a_1, b_1\}$, 
$\{a_1, b_2\}$, $\{a_2, b_1\}$, $\{a_2, b_2\}$. Each of these choices can be associated with a possible invariant set $I_{U_1}$, $I_{U_2}$, $I_{U_3}$ or $I_{U_4}$ respectively. Suppose, without loss of generality, Alice and Bob (independently) choose $a_1$ and $b_1$ respectively, implying that $\cos\theta_{a_1b_1}$ is describable by $N$ bits. By the cosine rule for spherical triangles
\be
\label{triangle}
\cos \theta_{a_2b_1}=\cos \theta_{a_1a_2} \cos \theta_{a_1b_1}+\sin \theta_{a_1a_2} \sin \theta_{a_1b_1} \cos \phi
\ee
where $0<\phi< \pi/2$ is the (typically small) angle subtended by the two sides of $\triangle_{a_1a_2b_1}$ at $a_1$ and $\phi/\pi$ is describable by $N$ bits (see Fig 3). Now the first term on the right hand side of (\ref{triangle}) is the product of two terms, each of which is describable by $N$ bits. Hence the product is base-2 rational. However, by a simple number theorem, $\cos \phi$ cannot be base-2 rational. Hence the left hand side of (\ref{triangle}) cannot be base-2 rational, and in particular cannot not describable by $N$ bits, no matter how big is $N$. The theorem is:

$\mathbf{Theorem}$\cite{Jahnel:2005}.  Let $\phi/\pi \in \mathbb{Q}$. Then $\cos \phi \notin \mathbb{Q}$ except when $\cos \phi =0, \pm 1/2, \pm 1$. 

$\mathbf{Proof}$. We derive a \emph{reductio ad absurdum}. Assume that $2\cos \phi = a/b$ is rational, where $a, b \in \mathbb{Z}, b \ne 0$ have no common factors.  Using the identity $2 \cos 2\phi = (2 \cos \phi)^2-2$ we have
\be
2\cos 2\phi = \frac{a^2-2b^2}{b^2}
\ee
Now $a^2-2b^2$ and $b^2$ have no common factors, since if $p$ were a prime number dividing both, then $p|b^2 \implies p|b$ and $p|(a^2-2b^2) \implies p|a$, a contradiction. Hence if $b \ne \pm1$, then the denominators in $2 \cos \phi, 2 \cos 2\phi, 2 \cos 4\phi, 2 \cos 8\phi \dots$ get bigger and bigger without limit. On the other hand, with $0 < \phi/\pi < 1/2 \in \mathbb{Q}$, then $\phi/\pi=m/n$ where $m, n \in \mathbb{Z}$ have no common factors. This implies that the sequence $(2\cos 2^k \phi)_{k \in \mathbb{N}}$ admits at most $n$ values. Hence we have a contradiction. Hence $b=\pm 1$ and $\cos \phi =0, \pm1/2, \pm1$. QED. 

Hence if in reality Alice chooses setting $a_1$ and Bob chooses setting $b_1$ when measuring the spins of a particle pair described by the hidden-variable $\lambda$, then the actual invariant set is $I_{U_1}$, and a counterfactual universe where Alice chooses setting $a_2$ and Bob chooses setting $b_1$ lies off $I_{U_1}$. Conversely, if in reality Alice chooses $a_2$ and Bob $b_1$, then the invariant set is $I_{U_3}$ and a counterfactual universe where Alice instead chooses $a_1$ and Bob $b_1$ lies off $I_{U_3}$. In general, only two of the four correlations in (\ref{chsh}) are definable for given $\lambda$, no matter what choices Alice and Bob make.

If only two of the four correlations are defined for any $\lambda$, what happens when experiments show that the inequalities are violated? The key point that distinguishes experimental procedure from the theoretical analysis above is that in an experimental test of the CHSH inequality the four correlation functions in (\ref{chsh}) are evaluated using four separate sub-experiments (implying four disjoint sets of hidden variables). According to Invariant Set Theory, each sub-experiment must be associated with a state of the universe on an invariant set $I_U$, i.e. the cosines of the relative angles must all be definable by $N$ bits. That is to say, according to Invariant Set Theory, what is actually tested experimentally is not (\ref{chsh}) but something of the form
\begin{equation}
\label{chsh2}
|\mathrm{Corr}_{\Lambda_1}(a_1, b_1) - \mathrm{Corr}_{\Lambda_2}(a'_1, b_2)|+|\mathrm{Corr}_{\Lambda_3}(a_2, b'_1)+\mathrm{Corr}_{\Lambda_4}(a'_2, b'_2)| \le 2
\end{equation}
where $\Lambda_1 \ne \Lambda_2 \ne\Lambda_3 \ne\Lambda_4$. Here $a'_1=a_1$, $a'_2=a_2$, $b'_1=b_1$ and $b'_2=b_2$ to within the necessarily finite precision of the measuring instruments, such that all of $\cos \theta_{a_1 b_1}$, $\cos \theta_{a'_1 b_2}$, $\cos \theta_{a_2 b'_1}$ and $\cos \theta_{a'_2 b'_2}$ are describable by $N$ bits. By contrast, the theoretical analysis described above, necessary to determine whether a given hidden-variable theory is constrained by the CHSH inequality, considers different measurement orientations for a given $\lambda$, a situation which can never arise in an experimental test of the CHSH inequality. 

In summary, the reason Invariant Set Theory can violate the CHSH inequalities is through a violation of the measurement independence condition (\ref{mip}), in particular that
\be
\label{mip}
\rho(\lambda| \hat a, \hat b)= \rho (\lambda| \hat a, \hat b')
\ee
where the cosine of the angle between $a$ and $b$ is not describable by $N$ bits (e.g. is not dyadic rational), but the cosine of the angle between $a$ and $ b' \approx b$ is. An immediate reaction to such a conclusion might be that such a violation would not be robust to the tiniest perturbation. We discuss this in Section \ref{fine} below, showing it is not so. 

It can be considered an open question as to whether all demonstrations of quantum non-locality are nullified in Invariant Set Theory by the incommensurateness of $\phi$ and $\cos \phi$. However, given the general applicability of such incommensurateness to the realistic interpretation of a range of quantum phenomena - interference, sequential Stern-Gerlach, and the original Bell inequality (see \cite{Palmer:2015b}) - we speculate here that this question can be answered in the positive. 

\section{Fine Tuning, Conspiracy, Free Will and Superdeterminism}
\label{conspiracy}

It is claimed above that the Bell Inequalities can be violated without resorting to a breakdown of realism or local causality. What are the conceivable objections to Invariant Set Theory as a description of physical reality? We list four: it is fine-tuned, inimical to experimenter free will, conspiratorial or superdeterministic. In this section it is shown that invariant set theory is none of these. 

\subsection{Fine Tuning}
\label{fine}

The analysis above suggests that it might be sensitive to tiny perturbations (experimenter hand shake for example). However, this is not the case. We return to one of the basic properties of a Cantor Set, that it is `large' (i.e. has the cardinality of the continuum) when looked at from the inside. That is to say, the analysis above is insensitive to perturbations which keeps a trajectory on the invariant set. There is a continuum of such perturbations. Indeed, one can perform continuous analysis on $I_U$ making use of the link to p-adic numbers. For example, there is a well-known continuous 1-1 mapping between the dyadic integers $...d_2d_1d_0$ and the points $0.e_0e_1e_2$ of the Cantor Ternary set, by the relationship $e_n=2d_n$ \cite{Robert}. 

On the other hand, the analysis is sensitive to noise which is random with respect to the measure of the euclidean space in which $I_U$ is embedded - such noise will surely take a state on $I_U$ to a state off it, no matter how small the noise actually is. However, since by the Cosmological Invariant Set Postulate, only those perturbations which maps points of $I_U$ to points of $I_U$ can be considered physical, then such full-measure noise is not physical. 

In this sense, the experimenter can certainly have a degree of hand shake. This will mean that parameters like $\theta$ in the discussion above are (epistemically) uncertain, and can be expected to vary during an experiment within some finite tolerance $\Delta \theta$. However, according to Invariant Set Theory, this uncertainty notwithstanding, the actual values of the parameter $\theta$ occurring within this finite tolerance must all be describable by $N$ bits. 

\subsection{Conspiracy}

Let us return to Bell's example (see Section \ref{introduction}) where the orientation of a measurement device is set by the output of a PRNG, itself sensitive to the value of the millionth digit of some input variable. Let us suppose the PRNG is run at $t=t_0$, and the actual experiment takes place at $t=t_1$. In principle (like the black-hole experiment), let us assume that $t_0 \ll t_1$. Now let us denote by $\mathcal A$ Bell's assertion (cf Section \ref{introduction}) that the value of the millionth digit is unimportant for any distinctly different purpose. Then, if $\mathcal A$ is true, there can be no rational reason to suppose the value of the millionth digit at $t_0$ should be correlated with the hidden variables associated with the particles whose spin is subsequently measured by Alice and Bob at $t_1$ (unless one invokes retrocausality).

However, assuming the Cosmological Invariant Set Postulate, there are three related ways of seeing that $\mathcal A$ is false. 

\begin{itemize}

\item
Consider an ensemble of putative states $X_U(d)$ of the Universe at $t_0$. By construction, suppose $X_U(d)$ are identical except for the value $d \in \{0,1,2, \ldots 9\}$ of the millionth decimal digit of the input variable to the PRNG. If $I_U$ is fractal, then it is impossible to determine by finite algorithm started at $t_0$, the value of $d$ which ensures $X_U(d) \in I_U$. The situation is fundamentally no different to that of the black-hole gedanken experiment. Similar to the discussion in Section \ref{cantor}, the value $d$ of the millionth digit for which $X_U(d) \in I_U$ is determined by events in the future of $t_0$, (a crucial) one of which will be the settings of the experiment to measure the spin of the particle pair at $t_1$. 

\item
Consider the implication of the Takens Embedding Theorem referred to in Section \ref{cantor}. In principle (but not in practice!) the entire invariant set $I_U$ could be reconstructed in state space, given sufficiently many values of the millionth digit of the input to the PRNG (over multiple aeons of the universe). This directly contradicts $\mathcal A$'s assertion that the values of the millionth digits are irrelevant for all distinctively different purposes. 

\item
Suppose the millionth digit was the number 8 and consider a counterfactual universe in which all degrees of freedom are kept fixed except for one, where the number 8 was perturbed to the number 9, taking $U \rightarrow U'$. Whilst $U \in I_U$, one cannot assume that $U' \in I_U$, since $I_U$ has measure zero in its embedding space. Above we have shown that at least some of the counterfactual measurement orientations needed to establish the CHSH inequality necessarily lie off $I_U$. Hence, states of the universe where the value of the millionth digit would have led to such counterfactual measurement orientations, rather than the actual measurement orientations, necessarily lie off $I_U$. 
\end{itemize}

\subsection{Free Will}

Alice and Bob can be said to have free will if they can hold rational beliefs that the future `is up to them'.  Holding a rational belief means that there is nothing that disprove such a belief. However, it doesn't mean that the rational belief is actually true. Here we take a `classical compatibilist' view \cite{kane} (whose proponents include notable philosophers such as Hobbes, Hume and Mill) that Alice and Bob are `free' when there is an absence of constraints or impediments preventing them from doing what they want to do. In the analysis of the CHSH experiment above, there are certainly no `knowable' constraints at the time when the experiment is performed, which prevent them from selecting any of the combinations $\{a_1, b_1\}$, $\{a_1, b_2\}$, $\{a_2, b_1\}$, $\{a_2, b_2\}$. This is consistent with the non-computability of $I_U$. Alice and Bob will makes their choices on the basis of what they want to measure, this notion of `want' being determined by actions of the neurons in their brains, processes which themselves act out on the invariant set. Having made that choice, say $\{a_1, b_1\}$, then two of the three remaining pairs of values can, retrospectively, be said to not lie on $I_U$. Hence the counterfactuals in which measurements along these off $I_U$ directions are not physical. 

In summary, according to the classical compatibilist view, the experimenters are free agents, determinism and the Cosmological Invariant Set Postulate notwithstanding. That is to say, they can hold rational beliefs that they really do `control' the settings (in the sense that they could have done otherwise), whereas in reality they do not (and, for some alternative settings, could not have done otherwise). 

\subsection{Superdeterminism}

The word `superdeterminism' is generally used to denote a class of theories in which the measurement independent condition \ref{mip} is violated. The word is generally used in a pejorative sense, with the implication that it such a theory is necessarily conspiratorial and denies free will. As shown above, Invariant Set Theory is not conspiratorial and does not deny free will. In this Section, we attempt a mathematical definition of the notion of superdeterminism and assess whether Invariant Set Theory can itself be considered superdeterministic.  

\begin{figure}
\centering
\includegraphics[scale=0.5]{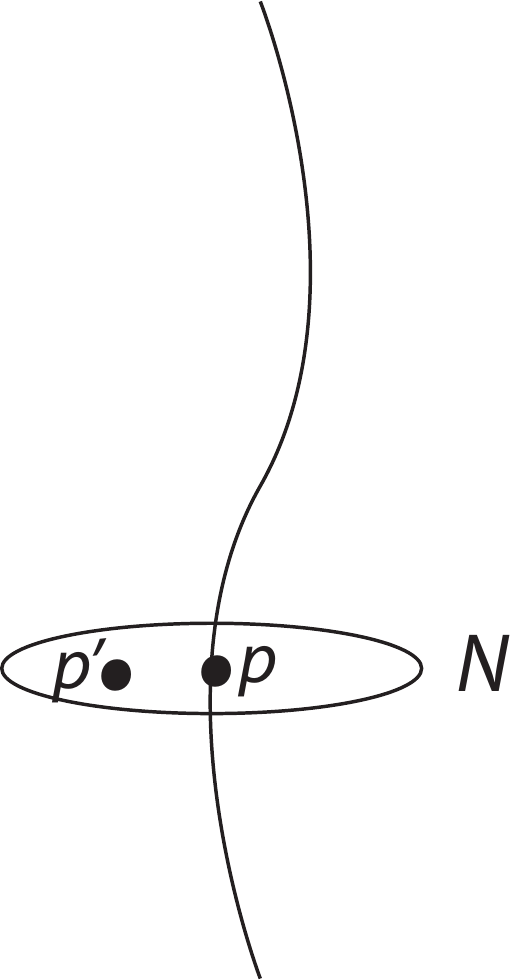}
\caption{Invariant Set Theory would be superdeterministic if for a point $p \in I_U$ there exist (sufficiently small) neighbourhoods $N$ of $p$, transverse to the state-space trajectory on $I_U$, which contain no other points of $I_U$. Because $I_U$ is locally the Cartesian product of the real line and a Cantor set, and because a Cantor set is a perfect topological set (with no isolated points), there exist (an infinity of) points where $p' \in N$ and $p' \in I_U$. Hence Invariant Set Theory is not superdeterministic.}
\label{fig:super}
\end{figure}

By definition, a superdeterministic theory is more restrictive than a deterministic theory - but how? A superdeterministic theory must not only have deterministic laws of evolution (e.g. in the form $\dot X=F[X]$),  it must also impose some restrictions on allowed initial states $X(t_0)$. Consider the set of state-space perturbations to $X(t_0)$, which are transverse to the (integral curve) trajectory of $\dot X=F[X]$ from $X(t_0)$ (see Fig \ref{fig:super}). If \emph{any} such perturbation to an allowed $X(t_0)$ produces an initial state forbidden by the theory, i.e. if $X(t_0)$ was a completely isolated point in all transverse neighbourhoods of $X(t_0)$, no matter how large, then of course it would be reasonable to say that such a theory is superdeterministic. However, this is an unnecessarily strong restriction and instead we will take a much weaker definition of superdeterminism: a deterministic theory is superdeterministic iff there exists a transverse neighbourhood of $X(t_0)$ (no matter how small) comprising only forbidden states. 

According to this weak definition, Invariant Set Theory is not superdeterministic. By the Cosmological Invariant Set postulate, states of the universe evolve on a fractal invariant set in state space - locally the product of a Cantor set and the real line. A Cantor set is a so-called `perfect' topological set. Any neighbourhood of a point of a perfect set contains other points of the set, i.e. a perfect set contains no isolated points. Hence all transverse neighbourhoods of an allowed initial state contain allowed initial states. 

Superdeterministic theories are rightly seen as unacceptable by physicists and philosophers. Most importantly, they offer no theoretical framework for defining the notion of probability (which is central to quantum theory). As such, they offer no theoretical framework to understand why the world is governed by quantum probability and not classical probability. This problem can be framed in the following way: if the world is indeed causal and deterministic (as a superdeterministic world could indeed be), how can we explain that the vast majority of quantum physicists have been fooled into believing otherwise. However as discussed, a fractal invariant set is not superdeterministic. In particular, any transverse neighbourhood of any state on $I_U$ contains an infinite number of states, also lying on $I_U$. From the statistics of these alternative states we can construct the notion of probability (and quantum probability in particular). 

\section{Conclusions}
\label{conclusions}

We have presented a new approach to formulating a locally causal hidden-variable theory of quantum physics, in which the measurement independence condition is partially violated without conspiracy, superdeterminism, retrocausality, fine-tuning and without denying experimenter free will. The key ingredient in this approach is the global invariant set; more specifically the notion that globally defined invariant subsets of state space define the notion of physical reality. Such subsets are generic in nonlinear dynamical systems theory, but are less familiar to quantum physicists and philosophers. To try to make accessible the reasoning why such subsets do provide a novel solution to the problem of quantum non-locality, we considered a simple black-hole gedanken experiment - the event horizon in space-time being the analogy of the invariant set in state space. The fundamental postulate underlying the analysis in this paper is the Cosmological Invariant Set Postulate, that states of the universe $U$ are evolving precisely on a global measure-zero fractal invariant set, $I_U$ in state space. 

Recently, a proposal was made to close the measurement independence `loophole' by setting measurement orientations from the light of distant quasars \cite{Gallicchio:2014}. Because the light originated so early in the universe, and because the cosmic sources have never been in causal contact with one another, it seems improbable (in line with Bell's reasoning about the PRNG) that this light could in any way be correlated with the hidden variables of particle source in an experiment being conducted today. However, by analogy with the black hole event horizon, it is shown that the cosmological invariant set is an atemporal one. The fact that the light was emitted billions of years before the experiment took place is really a `red herring'. It is therefore predicted that if these proposed experiments are conducted, Bell's inequalities will continue to be violated.

This experiment draws to attention the notion that if quantum theory can be replaced by a deeper causal deterministic theory, that theory must treat cosmology and quantum physics much more synergistically than is done in contemporary theories of physics. This in turn suggests that the long-sought unification of gravitational and quantum physics should also be based on the development of a more synergistic relationship between cosmology and quantum physics than exists today.

We conclude with some remarks about the implications of the discussions above on the physical nature of the central mathematical structure in quantum theory: the complex Hilbert space. A vector space is a rudimentary algebraic structure defined over a field, the field of complex numbers in the case of the complex Hilbert Space. The set of elements of a field is by definition closed over addition and multiplication. However, in terms of the correspondence (\ref{correspondence}),  $\cos \theta$ and $\phi/\pi$ are restricted to rational numbers describable by fixed finite $N$ bits. Hence, the set of corresponding numbers $\cos \theta$ is not in general closed under multiplication, and the set of corresponding numbers $e^{i\phi}$ is not in general closed under addition, no matter how large is $N$. Hence the set of vectors corresponding to bit strings describable in Invariant Set Theory is not a vector space, no matter how large is $N$. Is this a problem? Whilst it is appealing to be able to describe all elements of a physical theory by rudimentary mathematical structures, as a physicist one should not be overly beguiled by the elegance of any \emph{particular} mathematical structure. In particular, the property of closure guarantees that all conceivable counterfactual worlds have equal ontological status, the antithesis of that implied by the Cosmological Invariant Set Postulate. Hence we claim that closure can be an undesirable feature of theory if one strives for causal realism in physics. In particular, we reject the algebraic properties of the complex Hilbert Space as a description of physical reality and speculate that the unquestioning acceptance of these properties over the years has hindered the development of physical theory (the unification of quantum and gravitational physics in particular). In terms of Invariant Set Theory, the complex Hilbert space should be treated as the singular and not the regular limit of the set of bit string representations, as the parameter $1/N$ tends to zero. As Berry has noted \cite{berry}, old physical theories generically arise as the singular limit of new physical theories. 

\section*{Acknowledgement} My thanks to Harvey Brown, Huw Price, Terry Rudolph and Simon Saunders for helpful discussions.

\bibliography{mybibliography}

\end{document}